\documentclass[10pt, conference]{IEEEtran}

\usepackage[T1]{fontenc}
\usepackage[latin9]{inputenc}
\usepackage{float}
\usepackage{algorithm, algorithmicx, algpseudocode}
\usepackage{amsmath}
\usepackage{amssymb}
\usepackage{algorithm}
\usepackage{algpseudocode}
\usepackage{pifont}
\usepackage[none]{hyphenat}
\usepackage{url}
\usepackage{subfig}
\usepackage{multirow}
\usepackage{caption}
\usepackage[export]{adjustbox}
\usepackage{pgfkeys}
\usepackage{float}
\usepackage{booktabs}
\usepackage{balance}
\usepackage[scaled=.9]{beramono}

\usepackage{graphicx}
\newcommand{\fig}[1]{Fig.~\ref{#1}}
\newcommand{\figs}[2]{Fig.~\ref{#1}--\ref{#2}}
\newcommand{\figwidth}{0.875\columnwidth} 

\usepackage{xspace}

\newcommand*{\cf}{\textit{cf.}\@\xspace}
\newcommand*{\eg}{\textit{e.g.,}\@\xspace}
\newcommand*{\ie}{\textit{i.e.,}\@\xspace}
\makeatletter
\newcommand*{\etc}{%
    \@ifnextchar{.}%
        {etc}%
        {etc.\@\xspace}%
}
\newcommand*{\etal}{%
    \@ifnextchar{.}%
        {et al}%
        {et al.\@\xspace}%
}
\makeatother

\begin{document}
\title{Price and Performance of Cloud-hosted Virtual Network Functions: Analysis and Future Challenges}

\author{\IEEEauthorblockN{Nadir Ghrada\IEEEauthorrefmark{1}, Mohamed Faten Zhani\IEEEauthorrefmark{1}, Yehia Elkhatib\IEEEauthorrefmark{3}\IEEEauthorrefmark{1}}
\IEEEauthorblockA{\IEEEauthorrefmark{1}\'Ecole de Technologie Sup\'erieure (\'ETS), Montreal, Quebec, Canada\\
\IEEEauthorrefmark{3}MetaLab, School of Computing and Communications, Lancaster University, UK\\
E-mail: nadir.ghrada.1@ens.etsmtl.ca, mfzhani@etsmtl.ca, \{i.lastname\}@lancaster.ac.uk \\}
}  

\maketitle

\begin{abstract}
The concept of Network Function Virtualization (NFV) has been introduced as a new paradigm in the recent few years. NFV offers a number of benefits including significantly increased maintainability and reduced deployment overhead. Several works have been done to optimize deployment (also called \emph{embedding}) of virtual network functions (VNFs). However, no work to date has looked into optimizing the selection of cloud instances for a given VNF and its specific requirements. In~this~paper, we~evaluate the performance of VNFs when embedded on different Amazon EC2 cloud instances. Specifically, we evaluate three VNFs (firewall, IDS, and NAT) in terms of arrival packet rate, resources utilization, and packet loss. Our~results indicate that performance varies across instance types, departing from the intuition of ``you get what you pay for'' with cloud instances. We~also find out that CPU is the critical resource for the tested VNFs, although their peak packet processing capacities differ considerably from each other. Finally, based on the obtained results, we identify key research challenges related to~VNF instance selection and service chain provisioning.
\end{abstract}
\IEEEpeerreviewmaketitle
\section{Introduction}
Since its introduction as a concept by ETSI to~decouple software from hardware by leveraging virtualization technology \cite{ETSIWhitepaper}, Network Function Virtualization (NFV) has been widely and rapidly adopted as more cost-effective and easy to manage replacement to traditional hardware-based middleboxes. Such monolithic hardware solutions are~not just expensive to obtain and maintain, they also make it difficult to~re-allocate~network functions such as firewalls, load balancers, intrusion detection systems (IDS), and network address translation (NAT). In contrast, NFV enables such network functions to basically run on~virtual machines (VMs) hosted by commodity servers as~Virtual~Network~Functions~(VNFs).

NFV offers numerous benefits to cloud service providers (CSPs) including networking equipment cost reduction, power consumption minimization, scalability, elasticity, hardware reuse, easy multi-tenancy, and rapid configuration of new services \cite{ETSIWhitepaper}. 
NFV services are composed of a set of network functions traversing the path(s) from one or multiple sources to the destination. Each such composition of ordered VNFs is~referred to as a \emph{service function chain} (SFC). 

SFC embedding is an intensely active research area. However, little attention has been given to optimal selection of~cloud resources -- commonly the most popular infrastructure~-- for SFC embedding. This is a non-trivial challenge as several studies have identified that the `book value' of cloud instances as reported by CSPs are unreliable as performance indicators \cite{Ostermann2010,Samreen2016Daleel}.

In this paper, we tackle the problem of identifying the~optimal cloud instances for hosting VNF. We target Amazon EC2 as the market leading Infrastructure-as-a-Service (IaaS) provider. Through extensive experimentation, we~analyze the performance-price trade-off of four different EC2 instance types. 
Our investigation distinguishes the~problem into two main parts: instance selection and~dimensioning. More specifically, the paper aims to~answer the following questions:
\begin{enumerate}
    \item Which instance type best meets the performance requirements of the SFC?
    \item Which instance type provides the most cost-effective option for the constituent VNFs?
    \item How many of instances are needed per VNF?
\end{enumerate} 
We are also mindful of the varying requirements of different VNFs. As such, we use 3 different VNFs as the use cases of~our analysis, namely: firewall, IDS, and NAT.
\vspace{4pt}

The remainder of this paper is organized as follows. Section~\ref{sec:RelatedWork} reviews the state of the art. Section~\ref{sec:Analysis} presents the setup and results of our analysis, while Section~\ref{sec:disc} discusses the outcomes and their significance. This is followed by~Section~\ref{sec:future}, which focuses a few key future challenges in~light of our results, and Section~\ref{sec:Conclusion} concludes.

\section{Related Work}
\label{sec:RelatedWork}
\begin{figure*}[!t]
\centering
\includegraphics[width=0.90\textwidth]{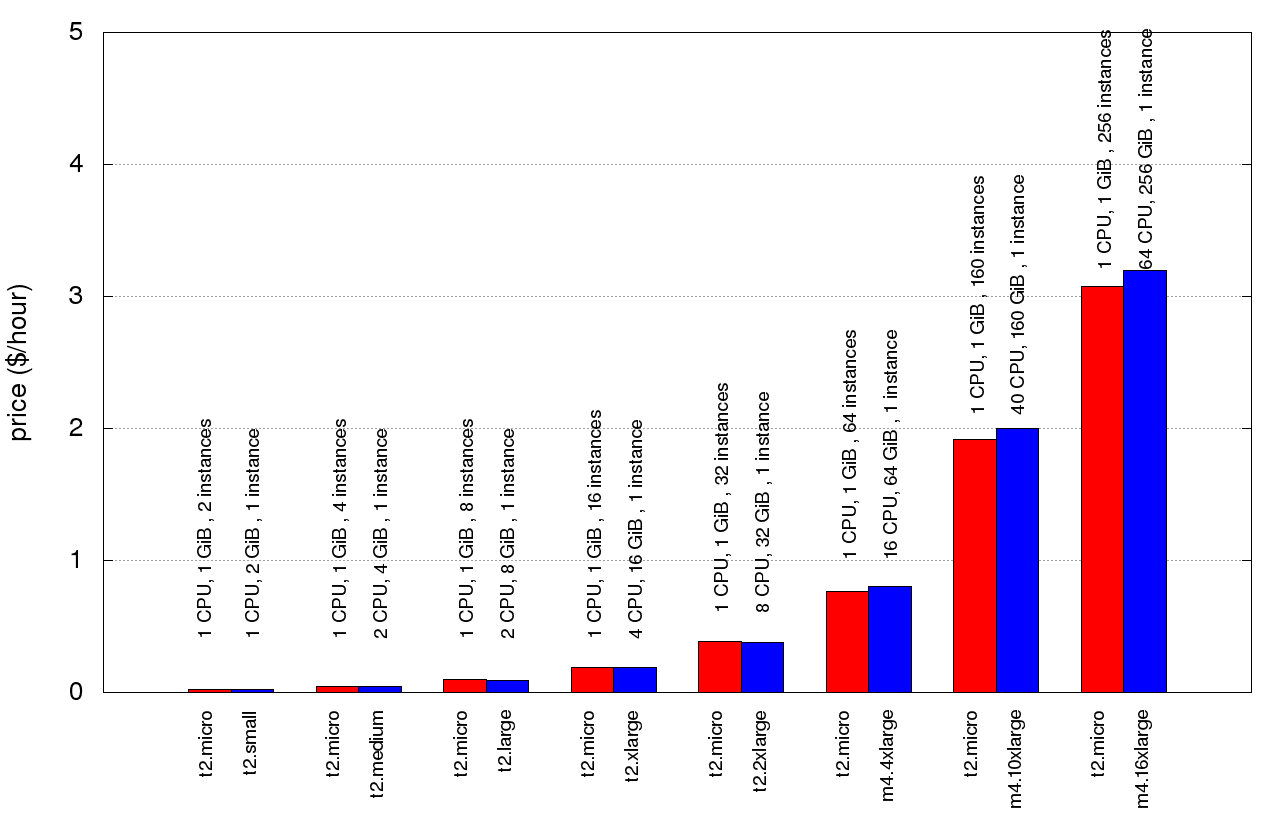} \label{fig:Amazon-EC2}%
\caption{Instance capacity versus Price}
\label{fig:vnf-capacity-price}
\end{figure*}

There is a growing body of work on VNF placement and chaining (\eg \cite{ElasticVNF2015,Racheg2017,carpio2017vnf}), and state management and~migration (\eg \cite{Kothandaraman2015,Peuster2016, ZhaniIM2013,ZHANI-IGI-Global13}). 
However, these approaches focus on theoretical optimization and overlook the~performance of the physical infrastructure performance.

Reliance on the specifications of cloud instances as reported by CSPs is severely problematic as such specifications are not necessarily reliable. An early study \cite{Ostermann2010} looked into variances between a few EC2 instance types using standard benchmarks and noted that there are no ``best performing instance type''. 
Other studies have identified variances within one provider \cite{5708447,Zhonghong2012,Li2013,7027861}, and over a span of several days \cite{6655674}, 
different times of the year \cite{Schad2010}, and different regions of~a~single IaaS provider \cite{lucas2013multi}. 
More recent efforts \cite{Samreen2016Daleel} indicate that cloud instance performance is difficult to foresee based on the information offered by CSPs (EC2 in particular). Thus,~selecting an optimal instance is a non-trivial decision.

Therefore, some have tried to gain better understanding of~the~potential performance of cloud instances using standard or bespoke benchmarking suites and various modelling techniques; \eg~\cite{6133223,6753834}. 
Others use profile-based methods (\cf~\cite{Cloudcmp,Verghese2013Cloudbench}) 
or application-specific performance models (\cf~\cite{Venkataraman2016Ernest}). 
A recent paper looked into the suitability of~containers for hosting VNFs \cite{8256024}. 
However, no work to date has identified how to optimize the choice of cloud instances for~VNFs, \ie ones that maximize service availability and~packet processing rate, and~minimize instance price. 

\section{Analysis of Instance Pricing and Performance}
\label{sec:Analysis}
In this section, we analyze the prices of Amazon EC2 instances versus their size in terms of~vCPU~and memory. We then study the performance of different types of VNF instances using such instances in terms of a number of~critical metrics: transmitted packets, packet arrival rate, CPU utilization and~packet loss rate. 

\subsection{Price versus amount of resources}

The cost of operating Amazon EC2 instances~\cite{AmazonEC2} is mainly reliant on their configuration (see Table~\ref{vnf-prices}), \ie the amount of virtual resources (\ie vCPU and memory) allocated to run these instances. 

\begin{table}[!th]
\centering
\caption{Excerpt of the configuration and price of the EC2 instances, as of November 2017.}
\label{vnf-prices}
\begin{tabular}{lrrr}
\toprule
\textbf{Instance Type} & \textbf{\begin{tabular}{@{}r@{}}vCPU \\ Cores\end{tabular}} & \textbf{\begin{tabular}{@{}r@{}}Memory \\ (GiB)\end{tabular}} & \textbf{\begin{tabular}{@{}r@{}}Price \\ (US\$/hour)\end{tabular}} \\ 
\midrule
\texttt{t2.micro} & 1 & 1 & 0.012 \\
\texttt{t2.small} & 1 & 2 & 0.023 \\
\texttt{t2.medium} & 2 & 4 & 0.047 \\
\texttt{t2.xlarge} & 4 & 16 & 0.188 \\ 
\bottomrule
\end{tabular}
\end{table}

\begin{figure*}[!tb]
\centering
\subfloat[Packet loss]{\includegraphics[width=\figwidth]{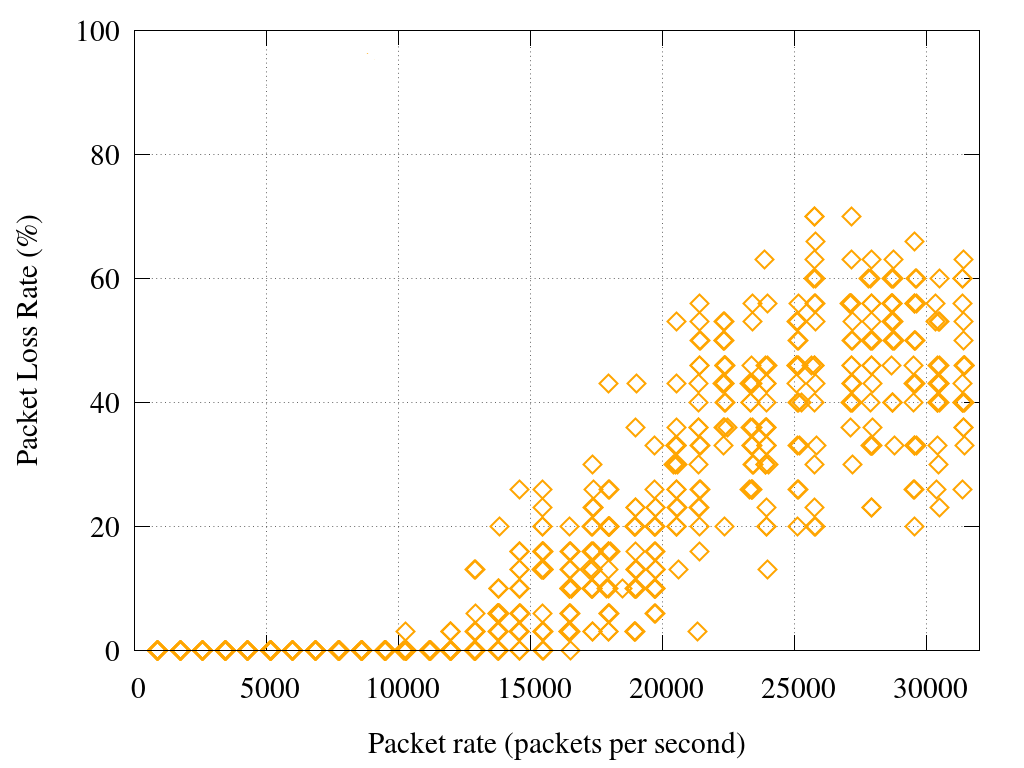}}\label{fig:pktloss-micro-fw}%
\subfloat[Resource utilization]{\includegraphics[width=\figwidth]{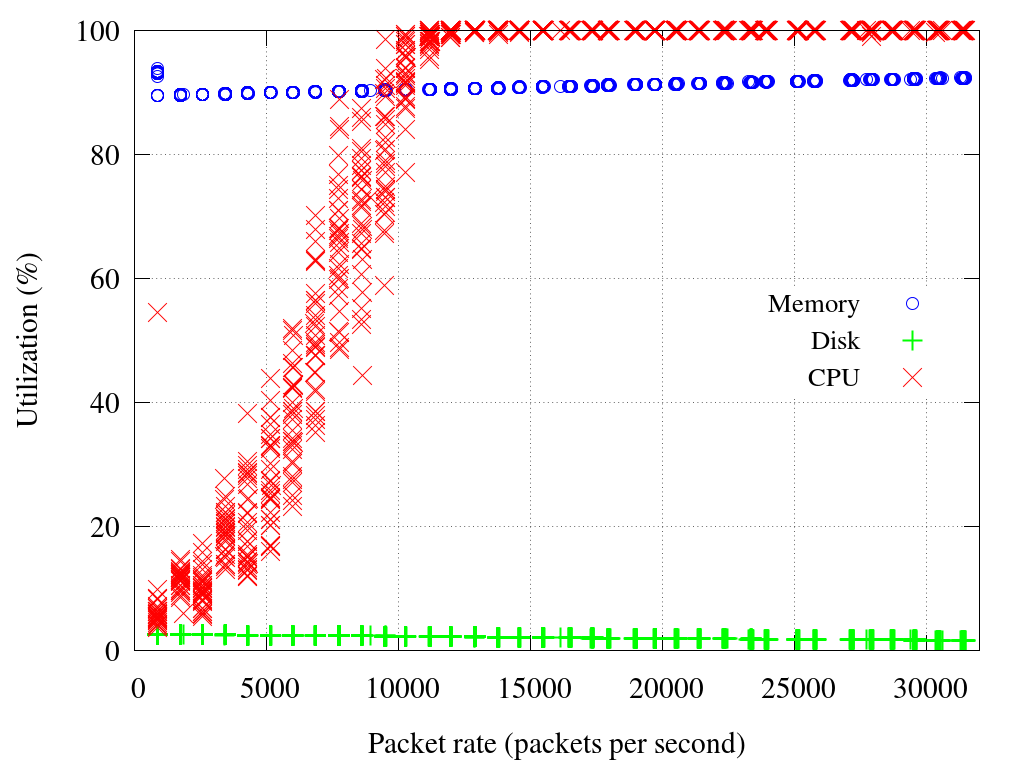}}\label{fig:stats-micro-fw}%
\caption{Firewall}\label{fig:resultsfirewall}
\end{figure*}

\begin{figure*}[!htb]
\centering
\subfloat[Packet loss]{\includegraphics[width=\figwidth]{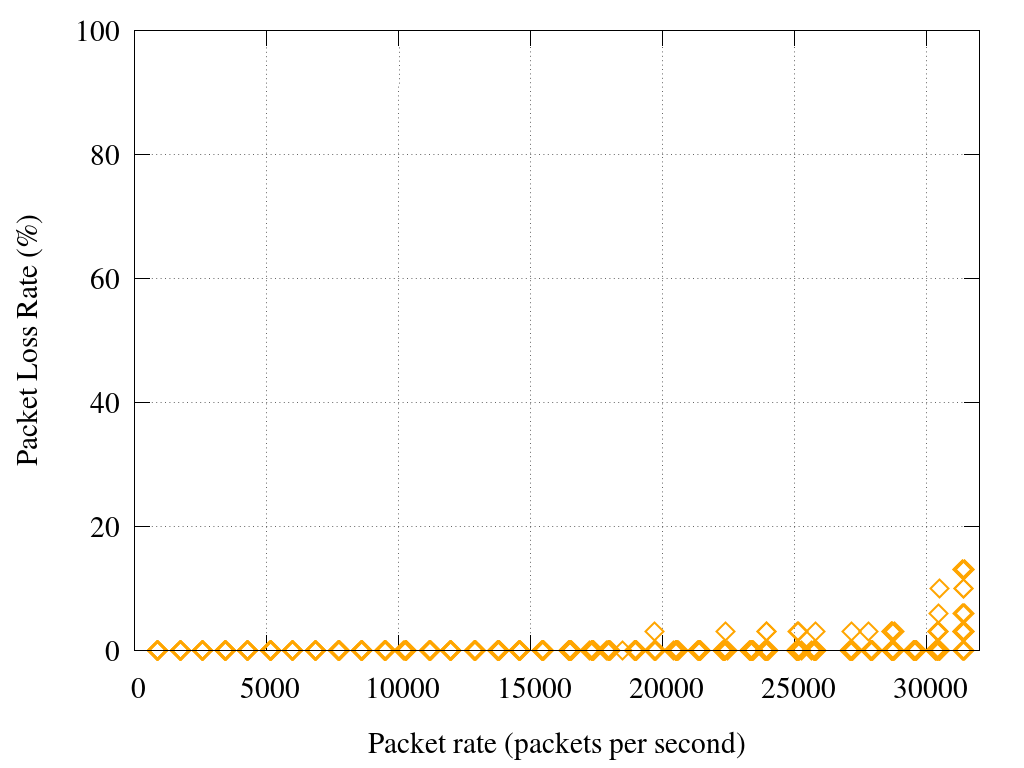}}\label{fig:pktloss-micro-ids}%
\subfloat[Resource utilization]{\includegraphics[width=\figwidth]{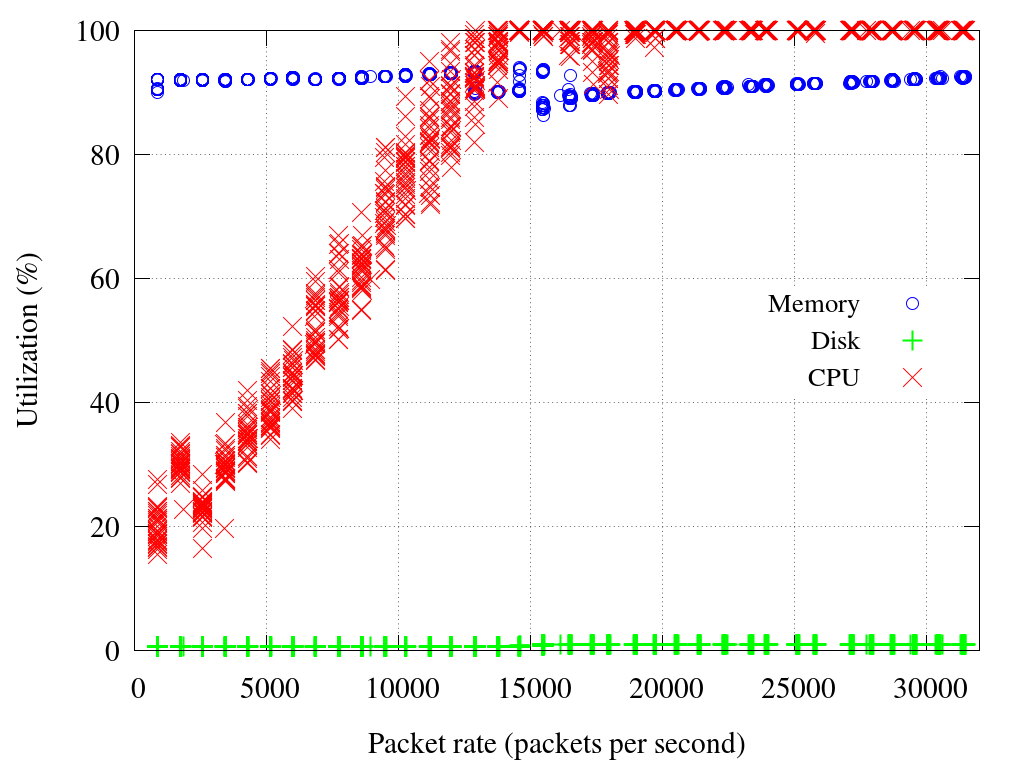}}\label{fig:stats-micro-fw}%
\caption{IDS}\label{fig:resultsIDS}
\end{figure*}

\begin{figure*}[!htb]
\centering
\subfloat[Packet loss]{\includegraphics[width=\figwidth]{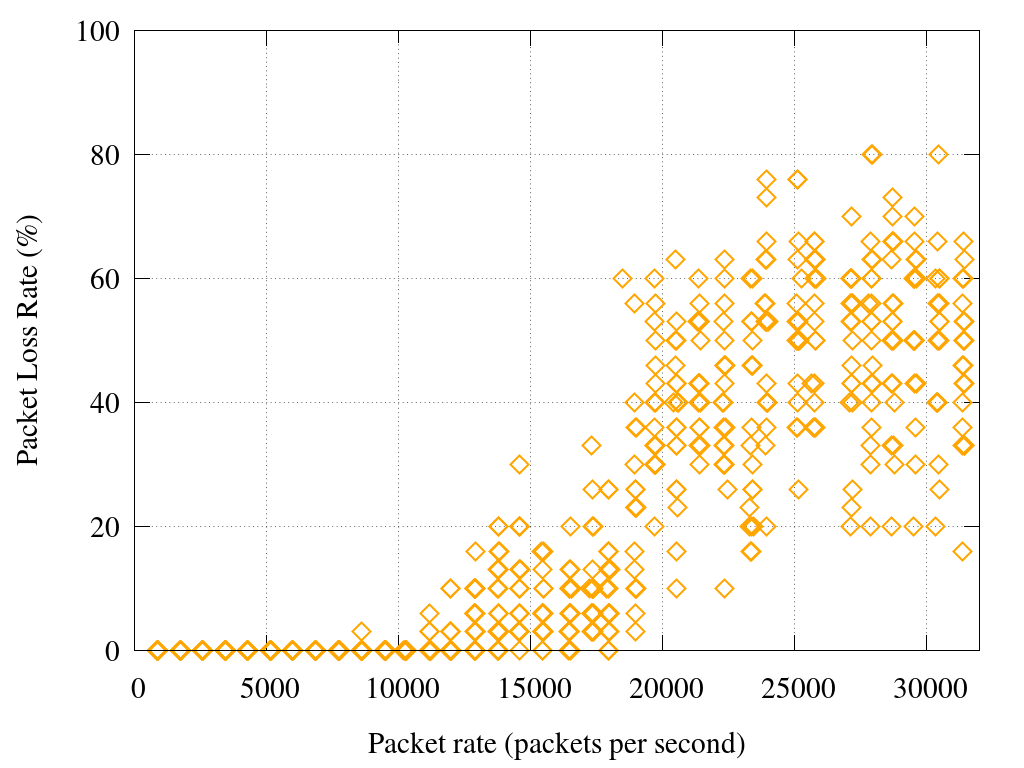}}\label{fig:pktloss-micro-nat}
\subfloat[Resource utilization]{\includegraphics[width=\figwidth]{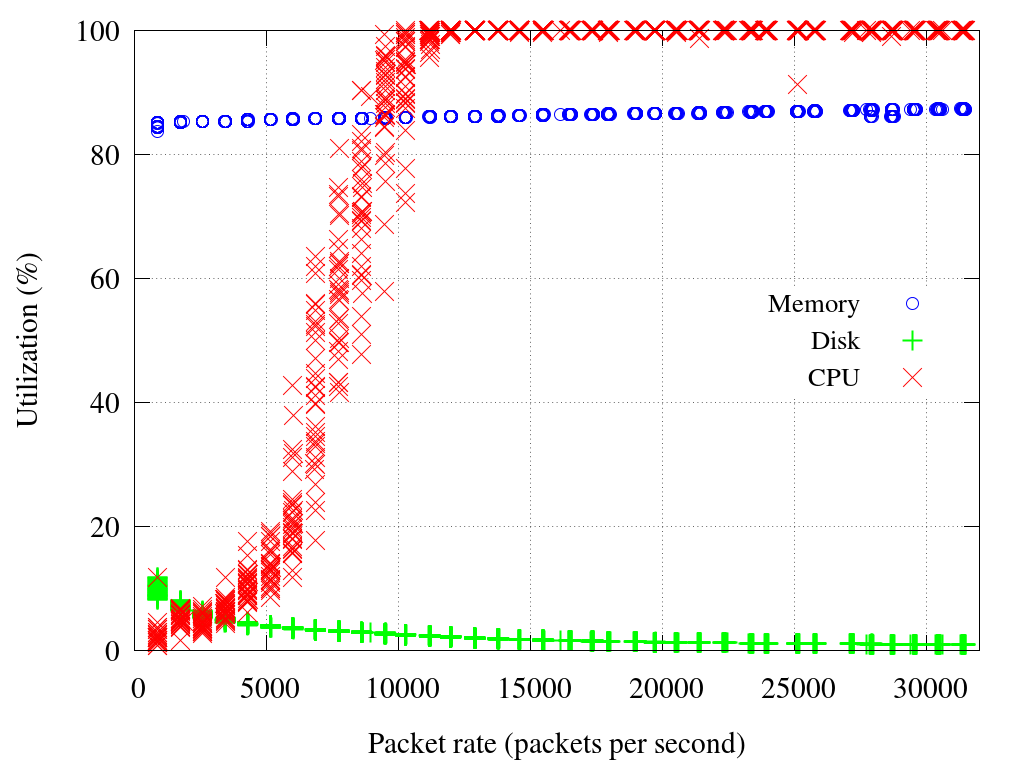}}
\caption{NAT}\label{fig:resultsNAT}
\label{t2.micro_1}
\end{figure*}

According to Amazon EC2 pricing list \cite{AmazonEC2}, the~lowest instance configuration is \texttt{t2.micro}, which operates with~the~least resources in terms of CPU and memory (\ie 1~vCPU core and~1~GiB of memory). The price to run a single instance of~\texttt{t2.micro}, for example, is \$0.012 per hour whereas operating an instance of \texttt{t2.small} type, which is allocated a single vCPU and 2GiB RAM, is~almost double the hourly rate at \$0.023. This also means that having two \texttt{t2.micro} instances would cost a tenant \$0.024 per hour. However, the tenant will have twice the~processing capacity of~a~single \texttt{t2.small}. In other words, two \texttt{t2.micro} instances would \emph{in theory} provide better performance than~a~single \texttt{t2.small} instance with merely the same price.

To better analyze whether this holds true for other types of~EC2~instances, we evaluate how many \texttt{t2.micro} instances it would take to provide the same amount of resources of larger EC2 instance types, and how do they compare in~terms~of~price. 

\fig{fig:vnf-capacity-price} shows the cost of each EC2 instance as well as the number and price of the \texttt{t2.micro} instances that would provide an equivalent amount of resources in terms of CPU and memory. For example, the largest EC2 instance, \texttt{m4.16xlarge}, provides 64 vCPU cores and~256~GiB, and~costs \$3.2/hour. The same amount of memory could be provided by 256~\texttt{t2.micro} instances with a~cost of~\$3.1/hour, but with significantly more vCPU cores (256~compared to 64). 
The same observation holds for all other types of~instances suggesting that using smaller instances (\ie~\texttt{t2.mirco}) would, theoretically, provide reduced costs and increased performance especially in terms of CPU horsepower.

\subsection{VNF performance versus instance type}

In this subsection, we evaluate the performance of three VNFs namely a firewall, an IDS, and a NAT. We ran Shorewall as a firewall \cite{Shorewall17}, Snort as a network IDS \cite{Snort17}, and the Ubuntu-based NAT. 

$\bullet$ \textbf{Experimental Setup:} 
We start the evaluation by generating UDP traffic between a pair of instances, where the VNF is installed inline to face the incoming traffic, process it, and then forward it to the destination. We ran the same experiment to examine the processing capacity of~all~VNFs.

We generated traffic using \texttt{iperf}, with this traffic increasing linearly with time. The purpose of increasing traffic load is to measure the CPU, memory, and disk utilization as well as packet loss for a VNF instance for different packet arrival rates. By doing this, we determine the highest packet processing capacity that each VNF can process per second in~a~particular instance.

$\bullet$ \textbf{Micro instance processing capacity:} 
As the previous subsection highlighted, the \texttt{t2.micro} instance offers the most cost-effective access to VNF-hosting cloud instances. Thus, we look at how capable such instance is at hosting VNFs.

For each VNF, we measure CPU, memory and disk utilization as well as packet loss as the received traffic increases over time. The results are depicted in \figs{fig:resultsfirewall}{fig:resultsNAT}. 
For~the firewall VNF running on a \texttt{t2.micro} instance, it~becomes overloaded and rejects packets from around 11,000 packets per second  (\fig{fig:resultsfirewall}(a)). This~corresponds to 90\% CPU and memory utilization (\fig{fig:resultsfirewall}(b)).

We also see in \fig{fig:resultsIDS} that an IDS running on~a~\texttt{t2.micro} instance behaves differently from a firewall VNF on the same instance. Packet losses start to occur when more than 18,000 packets start to arrive per second (\fig{fig:resultsIDS}(a)). This corresponds to 100\% and 90\% of CPU and~memory utilization (\fig{fig:resultsIDS}(b)), respectively.

Finally, the results for the NAT are reported in \fig{fig:resultsNAT}. We note that the performance of the \texttt{t2.micro}-hosted NAT starts to deteriorate around 9,000 packets per second arrival rate (\fig{fig:resultsNAT}(a)). This corresponds to 90\% CPU and memory utilization (\fig{fig:resultsNAT}(b)).

\textbf{$\bullet$ Amount of resources versus performance:} To compare the performance of instances of different sizes, we ran the same experiments for different instance types. For each of them, we identify the peak packet processing capacity (packets per second) as the maximum packet processing rate of the VNF when the CPU utilization reaches 90\% or the packet loss reaches 10\%, whichever first.

\begin{figure}[t]
\centering
\includegraphics[width=\figwidth]{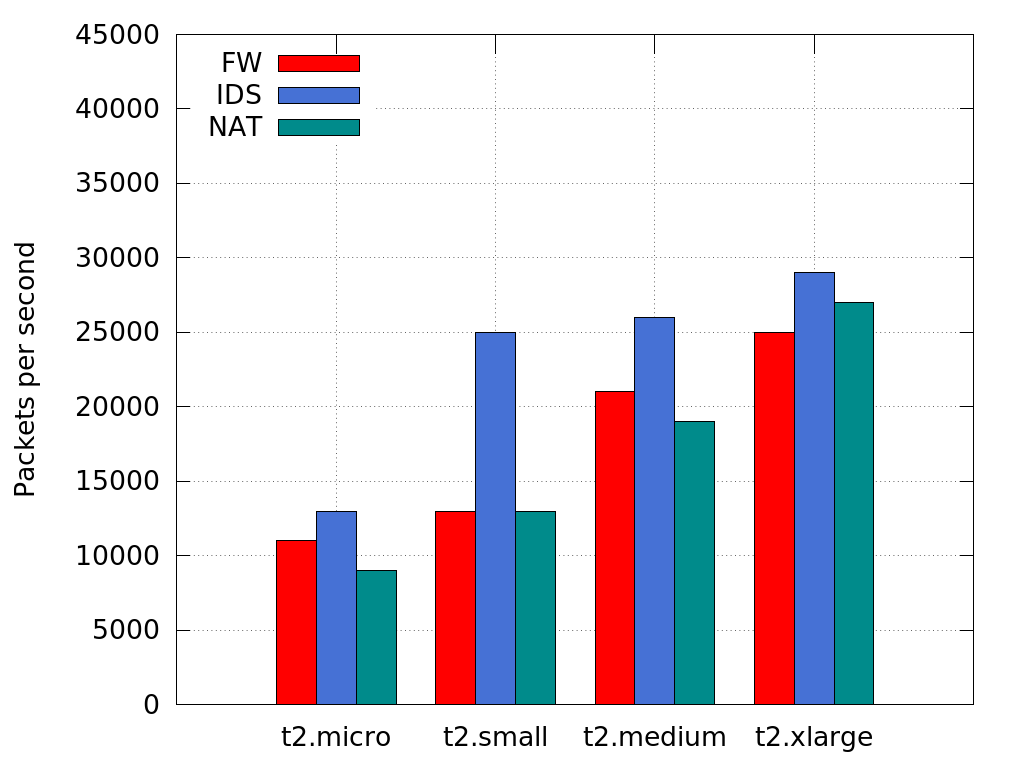} \label{fig:VNF-performance}%
\caption{VNF peak packet service rate per instance type, as dictated by CPU utilization of 90\% or packet loss of 10\%, whichever first}
\label{fig:vnf-instance-processing}
\end{figure}

\fig{fig:vnf-instance-processing}~summarizes the results and compares the packet processing capacity of four types of instances, \texttt{t2.micro}, \texttt{t2.small}, \texttt{t2.medium} and \texttt{t2.xlarge} for the three VNFs.
The first observation is that the~packet processing capacity may significantly vary depending on the type of the network function deployed in the instance. For example, the figure shows that the \texttt{t2.micro} can process as high as 13,000 packets/second when used to host a firewall, 13,000 packets/second for the IDS, and as low as 9,000 packets/second for the NAT.

We also notice that increasing the size of the instance results in a higher processing capacity. However, surprisingly, the increase in processing capacity is not proportional to the number of~CPU cores, nor to its price. For instance, int he case of a NAT (\fig{fig:vnf-instance-processing}), a \texttt{t2.micro} instance (1~CPU, 1~GiB, \$0.012/hour) can process 9,000 packets/second whereas a \texttt{t2.xlarge} (4~CPU, 16~GiB, \$0.188/hour) can process 26,000 packets/second; only an 188\% increase for more than 15x the price. In other words, even though \texttt{t2.xlarge} has 4 times more CPUs and 16 times more memory than \texttt{t2.micro}, it can process only 2.8 times more packets than~a \texttt{t2.micro} instance. This suggests that deploying several \texttt{t2.micro} instance provides better processing performance. 
Finally, recall that the price of 16 \texttt{t2.micro} instances is less than a single \texttt{t2.xlarge} (\fig{fig:vnf-capacity-price}), we can hence conclude that \texttt{t2.micro} instances are also more appealing in terms of~cost-effectiveness

\section{Discussion}
\label{sec:disc}
Abstracting from the above observations, we extract the following outcomes.

\subsection{Input packet rate is crucial; CPU is the bottleneck}
The incoming packet rate is the determining factor in the performance of VNFs. Consequently, anticipating such packet rate and its variance is a key element of dimensioning VNF embedding across any network. 

Based on the results plotted in \figs{fig:resultsfirewall}{fig:resultsNAT}, VNF packet processing rates could deteriorate due to increased contention on the virtual CPU resource. This is the case for all tested VNF types. Of course, an interesting avenue of future work would be to ascertain if this holds for a wider range of VNFs. 

\subsection{Not all VNFs are the same}
Every deployment of a VNF has its own performance `breaking point', which varies across VNF types and we conjecture that it would vary also between different implementations of the same VNF. We observed that a~low-demand VNF such as an IDS is easily hosted on~a~very modest cloud instance such as the EC2 \texttt{t2.micro}. However, to obtain similar packet rate capacity for a slightly more demanding VNF, such as a firewall, it~requires hosting on~a~higher specification instance (in this case, \texttt{t2.medium}). 

We can hence conclude that the type of the VNF and~its~implementation (in other words, the software running on the instance) will have a major impact on the packet processing capacity of the VNF. It is therefore a major challenge to write customized code that can optimize the~VNF operations and~further~leverage the resources available in~the~instance where the network function is running.

\subsection{Instance specification is a loose indicator of performance}
As described in Section~\ref{sec:RelatedWork}, this observation has already been made by different research groups using both generic benchmarks and specific application profiles. This has been a~motivator for our study, and indeed we have verified that such variance does exist even for NFV workloads. We found also the performance is~not~proportional to the amount of CPU and memory allocated for the VNF. This makes selecting the~right instance based on the amount of the instance resources not accurate enough. As~a~result, the instance specification is~only~a~loose indicator of~its~performance.
It is therefore of~utmost importance to test in advance the~VNF~when running on~a~particular instance, and~evaluate its~packet processing capacity as~it~may vary depending on~the~instance type. 

To conclude, we can say that current cloud provider offering models that provide generic VM instances with predefined resource capacity (\ie CPU, memory and disk) are not perfectly suitable for VNFs. A better offering model would be to provide VNFs with a well-defined network function software (software name and version) and an average packet processing capacity (in terms of packets per second).

\subsection{Micro instances are more cost-effective}

The obtained results clearly show that, compared to large instances, deploying multiple micro instances provides not only much more resources (in terms of CPU, memory and~disk), but also much more packet processing capacity with much less price. By combining results of \fig{fig:vnf-capacity-price} and \fig{fig:vnf-instance-processing}, we~can compare the performance, amount of resources and~prices for a single \texttt{t2.xlarge} instance and 16 \texttt{t2.micro} instances. As shown in Table~\ref{table:t2xlargeVSt2micro}, for the same price, we~can provision 16 \texttt{t2.micro} instances that provides 6 times the~packet processing capacity of a single \texttt{t2.xlarge}.

\begin{table}[]
\centering
\caption{Firewall performance using a single \texttt{t2.xlarge} instance compared to 16 \texttt{t2.micro} instances.}
\label{table:t2xlargeVSt2micro}
\begin{tabular}{lrrrr}
\toprule

  \textbf{\begin{tabular}[c]{@{}l@{}}Instance\\Type\end{tabular}} &  \textbf{\begin{tabular}{@{}r@{}}vCPU \\ Cores\end{tabular}} & \textbf{\begin{tabular}{@{}r@{}}Memory \\ (GiB)\end{tabular}}  &  \textbf{\begin{tabular}{@{}r@{}}Price \\ (US\$/hour)\end{tabular}}  & \textbf{\begin{tabular}[c]{@{}l@{}}Packet Processing \\ 
  Capacity (pckts/s)\end{tabular}} \\ 
  \midrule
\texttt{t2.xlarge} & 4 & 16 & 0.1856 & 26,500 \\ \midrule
\texttt{t2.micro}  & 16 & 16 & 0.1856 & 144,000  \\ \bottomrule
\end{tabular}
\end{table}

\section{Future Research Challenges}
\label{sec:future}
Based on these outcomes, we foresee future challenges in~this~domain to center around the following.

\subsection{VNF instance selection and placement}
Instance selection process is a daunting challenge that demands good knowledge of the VNF operational requirements and the performance of~the~hosting infrastructure \cite{Elkhatib2016crosscloudmap}. The~former requires analyzing the function operations, but the latter calls for a data-driven understanding of~the~performance of the underlying hosting infrastructure, and how this might vary over time. Indeed, hosting infrastructures are composed of several networking equipment, links and servers that are highly heterogeneous in capacity and performance. As a result, the performance of an instance with a predefined resource capacity might significantly vary in practice from one hosting server to~another and from one network to~another. Deciding of the instance type and its placement should therefore take into account the dynamic performance of the hosting equipment over time and space. The decision also depends heavily on the anticipated packet rate to be serviced. Therefore, a~central challenge related to~cloud-based VNF deployment is~to~put forward data-driven instance selection and placement solutions that are dynamic and adaptive.

A caveat here is that such data-driven approaches require significant amount of data to learn about the~hosting infrastructure performance levels and their variations. Such~data, obviously, comes at a cost, especially for~large-scale infrastructures. Consequently, further efforts are needed to~reduce such costs through either improved machine learning techniques (\eg \cite{Samreen2018tl}) or through data augmentation from external sources such~as~CloudHarmony~\cite{cloudharmony}.

\subsection{Dependency-aware service chain embedding}

In this work, we have only considered the performance-price tradeoff for hosting a single VNF in the cloud. This tradeoff naturally becomes significantly more complex for a SFC with more than one VNF as the decision for~any single VNF would be affected by that of other VNFs in the same chain. In this case, it is challenging to satisfy the  requirements of all VNFs in the chain while ensuring their dependency requirements (\eg packet rate, delay). 

Furthermore, our results suggest that micro-instances are more cost-effective than large ones. However, deploying the same function over several micro-instances is a non-trivial challenge as, depending on the type of the function and its requirements, some data might need to be shared among these instances. In this case, efficient synchronization techniques between the VNF instances need to be devised and evaluated to ensure normal operation of the overall network function. 
A typical example is that of an IDS. On one hand, when it is provisioned over a single large instance, there is no need for synchronization; however, the instance is expensive. On~the~other~hand, when several micro-instances of an IDS are deployed, they are able to process much more traffic (6x~according to~Table~\ref{table:t2xlargeVSt2micro}) at a cheaper price. The~instances analyze different shares of the incoming traffic so they need to pool information about detected attacks and incident analysis in order to make sure potential threats are detected. 

\subsection{Designing Serverless-based NFV-based service chains}
At a conceptual level, NFV lends itself to deployment using a serverless or Function-as-a-Service (FaaS) fashion. This is because many VNFs are relatively small, self-contained functions that are easily deployable across different locations in the network. More importantly, though, their operational cost depends heavily on the incoming service rate, as already highlighted. As such, they stand to benefit from FaaS deployment, which enables greater elasticity and flexibility in~the face of highly variable packet rates \cite{Baldini2017}. In this context, a major challenge would be to design serverless-based NFV service chains that are dynamically provisioned and torn down depending on the demand, the traffic routes, and the available resources in the underlying infrastructures.

\section{Conclusion}
\label{sec:Conclusion}
In this paper, we addressed the question of identifying the most suitable cloud instances for hosting VNFs, and the~effect of such decision on the VNF price and performance. We studied three different popular middlebox functions: firewall, IDS, and NAT. Our~experimentation spanned four different instance types from Amazon EC2. From~our~experiments, we surmise that na\"ive hosting based on instance specification alone is~suboptimal and costly. Multiple smaller instances provide much better performance-to-price ratio than a~single large instance. Based on our results, we discussed key research challenges to provide automated VNF instantiation and deployment solutions, and to investigate how this could affect service chain provisioning in cloud-based infrastructures.

\section*{Acknowledgment}
This work was partly supported by the Research Internationalization Fund at \'Ecole de Technologie Sup\'erieure (\'ETS) Montreal; and by the Adaptive Brokerage for the Cloud (ABC) project, UK EPSRC grant EP/R010889/1.

\balance
\bibliographystyle{IEEEtran}
\bibliography{RefBibtex}

\end{document}